# A conjecture on the origin of dark energy

Shan Gao
Unit for HPS & Centre for Time, SOPHI, University of Sydney
Email: sgao7319@uni.sydney.edu.au

The physical origin of holographic dark energy (HDE) is investigated. The main existing explanations, namely the UV/IR connection argument of Cohen et al, Thomas' bulk holography argument, and Ng's spacetime foam argument, are shown to be not satisfactory. A new explanation of the HDE model is then proposed based on the ideas of Thomas and Ng. It is suggested that the dark energy might originate from the quantum fluctuations of spacetime limited by the event horizon of the universe. Several potential problems of the explanation are also discussed.

Keywords: dark energy, holographic principle, holographic dark energy, UV/IR connection, bulk holography, spacetime foam, quantum fluctuations of spacetime

PACS: 95.36.+x., 04.20.-q, 98.80.-kx

## 1. Introduction

The origin of dark energy is an important issue in modern cosmology. Various solutions have been proposed to solve this great puzzle, one of which is the holographic dark energy (HDE) model [1-4]. According to the model, the dark energy density is

$$\rho_{DE} = 3d^2 M_P^2 L^{-2} \quad (1)$$

where $d$ is a numerical factor which is taken to be of the order of unity, $M_P$ is the reduced Planck mass $M_P^2 = 1/8\pi G$, $L$ is the event horizon of the universe. It has been shown that the HDE model is favored by the latest observational data including the sample of Type Ia supernovae (SNIa), the shift parameter of the cosmic microwave background (CMB), and the baryon acoustic oscillation (BAO) measurement (see, e.g. Refs. [5-6])[1]. However, a plausible physical explanation of the HDE model is still lacking [9-11]. For example, a recent analysis shows that the well-accepted explanation of Eq. (1), which is based on the UV/IR connection argument of Cohen *et al* [1], has serious drawbacks when applying the model to different eras of the universe [10-11][2]. In this paper, we will mainly investigate the physical basis of the HDE model.

The plan of this paper is as follows. In Section 2, we first examine Cohen *et al*'s argument based on energy bound [1]. If the energy bound is saturated, then the density of quantum zero-point energy assumes

---

[1] Note that the conclusions of Refs. [5-6] depend on the set of data used to constrain the HDE model. Moreover, in some HDE models where $L$ is not the event horizon of the universe (e.g. interacting HDE model [7]), the parameter $d$ may slowly vary with expansion in general [8].
[2] One conclusion of Ref. [11] is that "the basic framework underlying all HDE models seems too *ad hoc* to have any real

the same form as HDE in an effective quantum field theory (QFT) with the UV/IR connection required by the bound. However, it is shown that the theory cannot consistently describe all epochs of the universe and further explain the observed dark energy. As a result, Cohen *et al*'s argument is probably not the right physical explanation of the HDE model. In Section 3, we analyze Thomas' bulk holography argument based on entropy bound [2]. If the entropy bound is saturated, his method can also give the right form of the density of HDE. However, a concrete calculation shows that the method will give more vacuum energy than the observed dark energy. Therefore, it seems that the bulk holography argument cannot provide a plausible explanation of the HDE model either. These negative results suggest that the dark energy of the universe may not originate from the quantum zero-point energy in spacetime. In Section 4, we further examine the spacetime foam argument notably suggested by Ng, according to which the dark energy comes from the quantum fluctuations of spacetime [12-14]. It is shown that the argument also has several drawbacks. In particular, like Thomas' argument, it also predicts more energy than the observed dark energy. In Section 5, we propose a new model of HDE in terms of the quantum fluctuations of spacetime. It is shown that the model may not only give the right form of the density of HDE, but also be consistent with the observed dark energy. Conclusions are given in the last section.

## 2. The UV/IR connection argument of Cohen *et al*

The well-accepted explanation of the HDE model is that HDE comes from the quantum zero-point energy predicted by an effective QFT with a proper UV/IR connection. The argument was first given by Cohen *et al* to solve the fine-tuning problem of the cosmological constant [1], and it was then developed to explain the dark energy by Hsu and Li [3-4]. In the following, we will examine the argument in order to see whether it is the physical basis of the HDE model.

The argument of Cohen *et al* can be formulated as follows. For an effective QFT in a box of size $L$ with UV cutoff $\Lambda$, the entropy $S$ scales extensively, $S \sim L^3 \Lambda^3$. According to the holographic principle [15-17], the entropy $S$ should be limited by the Bekenstein-Hawking entropy bound, namely

$$L^3 \Lambda^3 \leq S_{BH} \sim L^2 M_P^2 \quad (2)$$

where $S_{BH}$ is the Bekenstein-Hawking entropy bound. Therefore, the length $L$, which acts as an IR cutoff, cannot be chosen independently of the UV cutoff and scales as $\Lambda^{-3}$. However, there is evidence that the above entropy bound is still loose, and in particular, a local QFT cannot be used as an effective low energy description of any system containing a black hole (e.g. particle states which size is smaller than their corresponding Schwarzschild radius) [16-17]. Therefore, there should exist a stronger constraint on the IR cutoff $L$, which excludes all states that lie within their Schwarzschild radius:

$$L^3 \Lambda^4 \leq L M_P^2 \quad (3)$$

where $\Lambda^4$ is the maximum energy density in the effective theory. Here the IR cutoff scales like $\Lambda^{-2}$.

---

explanatory value, which still keeps us in need of firmer theoretical background."

When Eq. (3) is near saturation, the entropy is $S_{\max} \approx S_{BH}^{3/4}$. Cohen *et al* suggested that an effective local QFT will be a good approximate description of physics when Eq. (3) is satisfied, because those states that cannot be described by it has been excluded. In other words, when the UV cutoff and the IR cutoff are properly connected, an effective local QFT will be still viable.

It is worth noting that Eq. (3) can also be derived by invoking the Bekenstein bound [10,18]. For a weakly gravitating system in which self-gravitation effects can be omitted, the Bekenstein bound is given by a product of the energy and the linear size of the system, $EL$. In the context of the effective QFT as described above, it is proportional to $L^4 \Lambda^4$. Then according to the holographic principle, we have $L^4 \Lambda^4 \leq S_{BH} \sim L^2 M_P^2$, and we can also obtain Eq. (3). Note that this requirement automatically prevents the formations of black holes, as the Bekenstein bound does not involve the Newton gravitational constant. Thus, the above two derivations are equivalent.

Now we analyze the validity of Eq. (3) for explaining the dark energy. Cohen *et al* argued that when choosing an IR cutoff comparable to the current horizon size of the universe, the corresponding UV cutoff obtained from Eq. (3) is about $10^{-2.5} ev$, and the resulting quantum energy density requires no cancellation and is consistent with current observations. Therefore, Eq. (3) can solve the fine-tuning problem of the cosmological constant. However, as first pointed out by Horvat et al [10-11], there may exist a loophole in Cohen *et al*'s derivation of the UV cutoff. According to the above UV/IR connection argument, an effective local QFT should be able to describe the standard models particles ($m \geq 100 Gev$) when Eq. (3) is satisfied. But when $\Lambda < m$ the energy density is not $\Lambda^4$ but $m\Lambda^3$, and thus we have $m\Lambda^3 \sim 10^{-10} ev^4$ and $\Lambda \sim 10^{-7} ev$. Consequently, the present-day UV cutoff is actually much smaller than $10^{-2.5} ev$ according to Eq. (3). As a result, the theory cannot describe the cosmic microwave background (CMB) radiation because the current temperature of the universe is $T_0 \sim 10^{-4} ev$ [11]. This inconsistency shows that the UV/IR connection argument based on Eq. (3) may have serious drawbacks when being used to explain the dark energy of the universe, and the dark energy may not originate from the quantum zero-point energy predicted by an effective QFT.

This conclusion has more support when applying Eq. (3) to other epochs of the universe. It has been argued that, when assuming most dark energy comes from the quantum zero-point energy satisfying Eq. (3), the matter-dominated epoch of the universe cannot be consistently described [10]. In order to solve this problem, some nonsaturated HDE models have been proposed. In these models, Eq. (3) is not saturated during the epochs that are not dominated by the dark energy. However, it is found that even such nonsaturated HDE models cannot account for the radiation-dominated epoch of the universe either [11]. The results are generic in that they do not depend on the choice of the IR cutoff. In conclusion, an effective QFT, which UV and IR cutoffs are connected by Eq. (3), cannot consistently describe all epochs of the universe, and thus it cannot explain the dark energy of the universe [11].

The above conclusion is also understandable by another analysis. When considering the success of the local QFT for describing the high-energy particles with a UV cutoff $\Lambda$ much larger than $10^{-2.5} ev$, the theory will be unable to consistently describe a very large system such as the whole universe, as the IR cutoff $L$ is much smaller than the size of the universe according to Eq. (3). Therefore, an inverse application of Eq. (3), namely using $L$ to limit $\Lambda$ as Cohen *et al* did, is probably improper when explaining the dark energy of the universe. In addition, there is another worry, namely that it may be problematic to take the left side of Eq. (3) as the quantum zero-point energy. There are some arguments against this direct equivalence. First, the energy is only predicted by an effective local QFT which eliminates those states that cannot be described by it. But such a theory is surely an incomplete description of actual situations. Moreover, the states that cannot be described by the theory do exist and may also have corresponding quantum zero-point energy. Obviously this part of energy is not included in Eq. (3). Next, the density of quantum zero-point energy in Eq. (3) is still local and extensive, which seems inconsistent with the spirit of the holographic principle, although the total energy satisfies a restriction. Besides, it is not obvious how to calculate the energy density in an effective QFT when the total energy is restricted. The left side of Eq. (3) implicitly assumes that the energy density integral is continuous from the IR cutoff to the UV cutoff. However, since the holographic principle requires that the number of degrees of freedom of any system is finite, it seems more natural that the integral is discrete and sparse in some sense, but still from the IR cutoff to the UV cutoff such as Planck's mass $M_P$. Lastly, the revision of the convention QFT must be radical due to the limitation of the holographic principle, and thus it is very likely that we should re-understand the quantum zero-point energy predicted by the conventional QFT. They may not exist in a fundamental theory (see, e.g. Refs. [19-20]).

To sum up, it seems that the dark energy of the universe cannot be accounted for by the quantum zero-point energy predicted by an effective QFT satisfying the UV/IR connection denoted by Eq. (3). Therefore, the popular explanation of the HDE model, i.e. that HDE comes from the quantum zero-point energy predicted by an effective QFT, is probably wrong.

### 3. Thomas' bulk holography argument

Another possible explanation of the HDE model is Thomas' bulk holography argument based on entropy bound [2] (see also [4,14]). The argument can be formulated as follows. In order to calculate a global quantum effect on the background geometry of the universe, it is natural to postulate that uniformly volume distributed bulk holographic degrees of freedom are delocalized on the scale of the background radius of curvature, denoted by *L*, since this is the relevant holographic length scale. The Heisenberg quantum energy of each delocalized holographic degree of freedom is $E \sim 1/L$. According to the holographic principle, the total number of the holographic degrees of freedom is $N \leq L^2 M_P^2$. Then the quantum contribution to the global vacuum energy density, $\rho_V \sim NE/L^3$, is:

$$\rho_V \leq M_P^2 L^{-2} \quad (4)$$

Such quantum contributions to the vacuum energy also satisfy the energy bound $NE \leq M_P^2 L$. Therefore,

holography allows only finite quantum corrections, and it provides a natural solution to the cosmological constant problem. This follows first from the holographic reduction in the number of independent degrees of freedom, and second from the holographic energy per degree of freedom.

It seems that Thomas' argument can also provide a plausible explanation of the HDE model when the holographic entropy bound is saturated. Let's analyze this claim in more detail. When the holographic entropy bound is saturated, the total number of the holographic degrees of freedom is $N \equiv A/4L_P^2 = \pi L^2 / L_P^2$, where $L$ is the horizon size of the universe, $A$ is the area of horizon, and $L_P$ is the Planck length. For the convenience of later analysis, we write down all parameters and constants explicitly. According to Thomas' argument, the Heisenberg quantum energy of each degree of freedom is $E \approx \frac{\hbar}{L} c = \frac{\hbar c}{L}$, where $c$ is the speed of light. Then the quantum contribution to the global vacuum energy density is:

$$\rho_V \approx \frac{NE}{4\pi L^3 / 3} = \frac{3c^4}{4GL^2} \quad (5)$$

If taking $L$ as the apparent horizon of the universe or the Hubble scale (i.e. $L = H^{-1}c$) as Thomas did [2], then the resulting energy density is obviously larger than the dark energy density. In fact, it is also larger than the critical energy density $\rho_c = 3H^2 c^2 / 8\pi G$. On the other hand, taking $L$ as the particle horizon cannot account for the accelerated expansion of the current universe (see, e.g. [4]). The promising alternative is taking $L$ as the event horizon of the universe. By using the definition of event horizon $L = a(t) \int_t^\infty dt' / a(t')$, we can solve the Friedmann equation for a spatially flat universe. The evolution equation of $\Omega_V$ is:

$$\frac{d\Omega_V}{d \ln a} = \Omega_V (1 - \Omega_V)(1 + \frac{2}{\sqrt{2\pi}} \sqrt{\Omega_V}) \quad (6)$$

where $\Omega_V \equiv \rho_V / \rho_c$. Then the equation of state up to the first order is:

$$w_V \approx -\frac{1}{3} \frac{d \ln \rho_V}{d \ln a} - 1 \quad (7)$$

By inputting the current value $\Omega_V \approx 0.72$, we obtain $w_0 \approx -\frac{1}{3}(1 + \frac{2}{\sqrt{2\pi}} \sqrt{\Omega_V}) \approx -0.56$. This result contradicts the latest observations of dark energy that requires $w_0 < -0.79$ (see, e.g. [21-22]).

We can also obtain the above negative result by directly invoking the observational restriction of the parameter $d$ in Eq. (1). Eq. (5) indicates $d = \sqrt{2\pi} \approx 2.5$. This value is too large to be able to explain the observed dark energy. For example, the latest best-fit results of Refs. [5] and [6] are respectively $d = 0.88^{+0.24}_{-0.06}$ and $d = 0.818^{+0.113}_{-0.097}$ for 68.3% confidence level. Considering the Heisenberg uncertainty principle, one may reduce the Heisenberg quantum energy to $E \approx \frac{\hbar/2}{L} c = \frac{\hbar c}{2L}$. Then the quantum contribution to the global vacuum energy density is:

$$\rho_V \approx \frac{NE}{4\pi L^3 / 3} = \frac{3c^4}{8GL^2} \quad (8)$$

This leads to $d = \sqrt{\pi} \approx 1.77$, which is still double of the best-fit value. Therefore, the saturated form of Eq. (4) cannot be consistent with the observational data of dark energy. Note that a holographic number of modes with the lowest frequency of quantum zero-point energy also gives more vacuum energy than the observed dark energy, as the quantum zero-point energy of the lowest frequency, $E_1 = \frac{hc}{8L}$, is still larger than the above Heisenberg quantum energy.

To sum up, the saturated form of Eq. (4) will lead to large dark energy density that is inconsistent with observations[3]. On the other hand, if the holographic entropy bound is not saturated, then Eq. (4), which is an inequality, cannot determine the concrete form of vacuum energy density alone, in particular, it will be unable to explain the $L^{-2}$ dependence of the HDE density. In conclusion, it seems that Thomas' bulk holography argument cannot provide a plausible explanation of the HDE model either. But it might give a clue to the right explanation, as there is only a numerical factor ~1/4 missed in the vacuum energy density formula Eq. (8).

## 4. Ng's spacetime foam argument

The failure of the arguments of Cohen *et al* and Thomas may reveal something positive about the nature of dark energy. It is that the dark energy of the universe may not originate from the quantum zero-point energy. On the other hand, it has been widely argued that spacetime as a dynamical entity should have quantum fluctuations (see, e.g. [12-13, 23-24]). Therefore, the quantum fluctuations of spacetime will contribute to the vacuum energy, and it may be the origin of dark energy. In short, dark energy might come from quantum fluctuations of spacetime, not from quantum fluctuations in spacetime.

According to Ng [12-13], spacetime, like all matter and energy, undergoes quantum fluctuations, and these quantum fluctuations make spacetime foamy on small spacetime scales. In order to know how foamy spacetime is, one needs to measure spacetime. By analyzing a Gedanken experiment to measure distance between two points, which was first suggested by Wigner [25-26], Ng concluded that the uncertainty $\delta L$ in the measurement of the distance $L$ cannot be smaller than the cube root of $LL_P^2$, namely $\delta L \geq L_P^{2/3} L^{1/3}$. Quantum mechanics requires $\delta L^2 \geq \frac{\hbar L}{mc}$, and general relativity requires $\delta L \geq \frac{Gm}{c^2}$,

---

[3] Note that this conclusion may also hold true for the models including interactions between dark energy and the matter sector. The reason is that when the universe is dominated by the dark energy, the energy density given by the saturated form of Eq. (4) is still larger than the critical energy density.

where $m$ is the mass of the clock used in the distance measurement. The product of these two inequalities then yields the above result. Similarly, the uncertainty $\delta T$ in the measurement of a time interval $T$ cannot be smaller than the cube root of $TT_P^2$, namely $\delta T \geq T_P^{2/3} T^{1/3}$, where $T_P$ is the Planck time. These results were also obtained by Károlyházy et al from somewhat different arguments [27-28].

The above spacetime uncertainty relation is consistent with the holographic entropy bound $S_\Lambda = \Lambda^3 L^3 \leq S_{BH} = M_P^2 L^2$ when the relation between the UV cutoff and distance uncertainty is $\Lambda \sim \frac{1}{\delta L}$ [4] [14]. By assuming each minimum detectable space cube $(\delta L)^3 \sim L_P^2 L$ has typical Heisenberg energy of a delocalized state $E \sim 1/L$, the energy density of the quantum fluctuations of spacetime is $\rho_V = \frac{E}{(\delta L)^3} \sim M_P^2 L^{-2}$, and it assumes the same form as the HDE density denoted by Eq. (1) [14]. A similar result is also obtained by Maziashvili in terms of time uncertainty [29], and it leads to the agegraphic dark energy model where the age of the universe determines $L$ [30].

The spacetime foam argument seems to provide a plausible explanation of HDE. However, it also has some potential problems. First of all, it is still in debate whether the quantum fluctuations of spacetime assume the very form $\delta L \sim L_P^{2/3} L^{1/3}$. Some authors have argued that the derivation of Ng is problematic [31-32], and different forms of spacetime fluctuations such as $\delta L \sim L_P^{1/2} L^{1/2}$ have also been suggested [33-34]. Next, even if Ng's derivation of the minimum distance uncertainty in a Gedanken measurement is valid, it does not necessarily entail that spacetime itself does have the similar uncertainty or fluctuations. Maybe it is only that the physical principles lead to an intrinsic limitation to spacetime measurements. Lastly, if the quantum fluctuations of spacetime indeed assume the very form suggested by Ng, then the holographic energy density will have the same form as Eq. (8), namely $\rho_V \approx \frac{3c^4}{8GL^2}$, as the calculation is the same as that in Thomas' method (see also [14]). However, as we have shown in the last section, this energy density is about the quadruple of the observed dark energy density.

In conclusion, although the spacetime foam argument may not provide a satisfactory explanation of HDE, it does suggest a promising possibility, namely that the holographic dark energy may come from quantum fluctuations of spacetime, not from quantum fluctuations in spacetime.

## 5. A conjecture on the origin of dark energy

In this section, we will show that a proper revision of Thomas and Ng's ideas may provide a possible explanation of the HDE model, and it is also consistent with the latest observation data of dark energy (see also [35-36]).

Following Ng's spacetime foam argument, we also assume that the holographic dark energy comes from the quantum fluctuations of spacetime. Following Thomas' bulk holography argument, we further assume each degree of freedom of such quantum fluctuations is also delocalized. But different from both of these arguments, we assume that the degrees of freedom are delocalized on the scale of the event horizon of the universe. In other words, we assume that the universe is a finite system limited by its event horizon in space, and the dark energy comes from the quantum fluctuations of the spacetime limited in the event horizon. This assumption has two interesting consequences. First, the Heisenberg quantum energy of one degree of freedom will be $\varepsilon \approx \frac{\hbar/2}{2L}c = \frac{\hbar c}{4L}$. Note that the space size limited by the event horizon is 2*L*, not *L*. This is equivalent to introducing one numerical factor 1/2 into Eq. (8) in Thomas' model. Next, since such quantum fluctuations of spacetime of one degree of freedom corresponds to two Planck area units at the two ends of the event horizon, the total number of degrees of freedom for such quantum fluctuations is $N/2 = \pi L^2 / 2L_P^2$. Note that the holographic principle implies that the event horizon contains finite area units, whose number is $N \equiv A/4L_P^2 = \pi L^2 / L_P^2$. This is equivalent to introducing another numerical factor 1/2 into Eq. (8) in Thomas' model. Therefore, the energy density of the quantum fluctuations of spacetime limited by the event horizon of the universe is:

$$\rho_V \approx \frac{\varepsilon N/2}{4\pi L^3/3} = \frac{3c^4}{32GL^2} \quad (9)$$

---
[4] This relation seems reasonable because the UV cutoff usually determines the minimal detectable length.

Compared with Eq. (8) in Thomas' model, Eq. (10) gains an additional numerical factor 1/4. This additional factor comes not from a mathematical trick, but from a different physical explanation. Eq. (10) indicates $d \approx \sqrt{\pi}/2 \approx 0.886$. This value is consistent with the latest observations [5-6].

In the following we will give several comments on this new explanation of the HDE model. First, it should be stressed that the physical nature and precise mathematical description of the quantum fluctuations of spacetime are still unknown, as a complete theory of quantum gravity is not yet available. However, it has been widely argued that spacetime should undergo some kind of quantum fluctuations, and they at least include the fluctuations of spacetime metric (see, e.g. [12-13, 23-24]). Despite these uncertainties, the above model might be also applicable because it only depends on the total number of degrees of freedom of such fluctuations and the fluctuation energy of each degree of freedom.

Secondly, the choice of event horizon in our model may have a physical basis. Contrary to the apparent horizon, the event horizon represents a real boundary of spacetime, and thus the quantum fluctuations of spacetime should be limited by the event horizon, not by other horizons. Moreover, the event horizon in the context of cosmology as well as in the context of a black hole is always defined globally, as the causal structure of spacetime is a global thing (see more discussions in [4]). Here one may raise a circularity problem (see also [9]). The HDE needs a finite event horizon, while a finite event horizon also needs HDE (without a dark energy or a cosmological constant to induce acceleration, the event horizon is necessarily infinite). Then what is first, HDE or event horizon? As we think, this is indeed a potential problem. However, it is not completely unsolvable at least. For example, the existence of both HDE and event horizon may be the results of the complete evolution law of the universe with certain initial condition, and there is no question of which is first. A universe without dark energy and event horizon is likely to exist too. Besides, there may also exist a small cosmological constant or other forms of exotic matter to induce the acceleration of the universe, and thus a finite event horizon may always exist. No doubt, further study is needed to solve this issue in a more satisfactory way.

Thirdly, we stress that the use of Heisenberg's uncertainty principle for spacetime fluctuations is still a tentative assumption, and it needs to be further justified. As we think, it might be reasonable to assume that any physical entity, no matter it is a matter field or a gravitational field, will have quantum fluctuations when limited in a finite space interval, and the fluctuation energy also satisfies Heisenberg's uncertainty principle. This assumption is also used to derive the dark energy density in Thomas and Ng's models [2, 14, 29][5]. As a result, the energy is only related to the spatial scale, and especially, it is irrelevant to the nature of the field. For example, for a gravitational field the fluctuation energy of one degree of freedom does not contain the gravitational constant $G$. However, the total fluctuation energy in a finite region contains $G$ as indicated by Eq. (10). Besides, it is worth noting that the uncertainty relations for the length and time fluctuations of a spacetime region may directly contain the gravitational constant through the involved Planck scale (see, e.g. [12, 33-34]). Certainly, whether this assumption is right or not can only be determined by experiments.

To sum up, the above physical explanation of the HDE model seems tenable. Moreover, it may help to solve some problems plagued by the HDE model, e.g. the IR cutoff choice problem, the saturated/ unsaturated problem and so on. In addition, the analysis also implies that the dark energy of the universe may originate from the quantum fluctuations of spacetime limited by its event horizon.

## 6. Conclusions

It is generally considered that the holographic dark energy (HDE) comes from the quantum zero-point energy predicted by an effective QFT with the UV/IR connection suggested by Cohen *et al*. However, it has been pointed out by Horvat that such a theory cannot consistently describe all epochs of the universe. Moreover, the UV/IR connection argument based on energy bound also has some serious drawbacks. Therefore, the well-accepted explanation of the HDE model is probably wrong. Different from the UV/IR connection argument, Thomas presented a bulk holography argument based on entropy bound, which has been regarded as another support for the HDE model. Although his method can give the right form of the density of HDE when the entropy bound is saturated, a concrete calculation shows that it will give more vacuum energy than the observed dark energy. Thus it seems that the bulk holography argument cannot provide a plausible explanation of the HDE model either.

The failure of the arguments of Cohen *et al* and Thomas may reveal something positive about the nature of dark energy. Maybe the dark energy of the universe does not originate from the usual quantum zero-point energy. Ng's spacetime foam argument is an important attempt along this line of thinking, according to which the dark energy comes from a special form of quantum fluctuations of spacetime.

---

[5] Note that Thomas seemed to also implicitly use this assumption because the holographic vacuum energy in his argument

However, this argument also has several drawbacks. In particular, like Thomas' argument, it also predicts more energy than the observed dark energy.

Inspired by the ideas of Thomas and Ng, we further propose a new explanation of the HDE model. It is suggested that the dark energy of the universe may originate from the quantum fluctuations of spacetime limited by the event horizon of the universe. By using the holographic principle and Heisenberg's uncertainty principle, it is shown that the energy density of such fluctuations assumes the same form as Eq. (1) in the HDE model. Moreover, the value of the numerical constant in Eq. (1), which turns out to be $d \approx \sqrt{\pi}/2$, is also consistent with the latest observations. Therefore, our proposal might provide a plausible physical explanation of the HDE model. Besides, it also suggests that the dark energy may come from the quantum fluctuations of spacetime, not from the quantum fluctuations in spacetime such as quantum zero-point energy.

**Acknowledgments** I am very grateful to the anonymous referees for his very insightful and helpful comments, suggestions and encouragement.

---

may include the contributions from the quantum fluctuation of the gravitational field [2].